\documentclass[prd,amsmath,amsfonts,floatfix,nofootinbib,preprintnumbers,superscriptaddress,showpacs]{revtex4}
\usepackage{amsfonts}
\usepackage{graphicx}
\usepackage{amsmath}
\usepackage[usenames]{color}

\newcommand*{\chpt}{\raise0.4ex\hbox{$\chi$}PT}
\newcommand{\negcdot}{\negmedspace\cdot\negmedspace}

\newcommand{\Delstar}{\ensuremath{\Delta^{\raise0.18ex\hbox{${\scriptstyle *}$}}}}
\def\gtwid{{\,\raise.35ex\hbox{$>$\kern-.75em\lower1ex\hbox{$\sim$}}\,}}
\def\ltwid{{\,\raise.35ex\hbox{$<$\kern-.75em\lower1ex\hbox{$\sim$}}\,}}
\def\leftvec{{\raise1.5ex\hbox{$\leftarrow$}\kern-1.00em}}
\def\rightvec{{\raise1.5ex\hbox{$\rightarrow$}\kern-1.00em}}
\def\half{{\scriptstyle \raise.2ex\hbox{${1\over2}$}}}
\def\threehalves{{\scriptstyle \raise.15ex\hbox{${3\over2}$}}}
\def\third{{\scriptstyle \raise.15ex\hbox{${1\over3}$}}}
\def\third{{\scriptstyle \raise.15ex\hbox{${1\over3}$}}}
\def\twothirds{{\scriptstyle \raise.15ex\hbox{${2\over3}$}}}
\def\fourth{{\scriptstyle \raise.15ex\hbox{${1\over4}$}}}
\def\la{\langle}
\def\ra{\rangle}

\def\op{{\mathcal{O}}}
\def\ampl{{\mathcal{M}}}

\def\exponential{{\mathrm{e}}}

\def\calP{{\mathcal{P}}}

\def\calV{{\mathcal{V}}}

\def\beq{\begin{equation}}
\def\eeq{\end{equation}}
\def\barr{\begin{array}}
\def\earr{\end{array}}

\newcommand{\vslash}{\ensuremath{v\!\!\! /}}

\newcommand{\cL}{\ensuremath{\mathcal{L}}}
\newcommand{\cM}{\ensuremath{\mathcal{M}}}

\newcommand{\cO}{\ensuremath{\mathcal{O}}}
\newcommand{\cP}{\ensuremath{\mathcal{P}}}

\newcommand{\cV}{\ensuremath{\mathcal{V}}}

\newcommand*{\Tr}{\operatorname{Tr}}

\newcommand{\bbar}{\ensuremath{\overline{b}}}
\newcommand{\qbar}{\ensuremath{\overline{q}}}
\newcommand{\Qbar}{\ensuremath{\overline{Q}}}
\newcommand{\Dbar}{\ensuremath{\overline{D}}}
\newcommand{\Bbar}{\ensuremath{\overline{B}}}
\newcommand{\cbar}{\ensuremath{\overline{c}}}
\newcommand{\ubar}{\ensuremath{\overline{u}}}

\newcommand{\Hbar}{\ensuremath{\overline{H}}}

\newcommand*{\bea}{\begin{eqnarray}}
\newcommand*{\eea}{\end{eqnarray}}
\newcommand*{\be}{\begin{equation}}
\newcommand*{\ee}{\end{equation}}

\begin{document}

\title{Possible lattice approach to $B\to D\pi (K)$ matrix elements}
\author{Christopher~Aubin}
\email[]{caubin@fordham.edu}
\affiliation{Department of Physics, Fordham University, Bronx, NY 10458, USA}
\author{C.-J.~David~Lin}
\email[]{dlin@mail.nctu.edu.tw}
\affiliation{Institute of Physics, National Chiao-Tung University, Hsinchu 300, Taiwan}
\affiliation{Division of Physics, National Centre for Theoretical Sciences, Hsinchu 300, Taiwan}
\author{Amarjit~Soni}
\email[]{soni@quark.phy.bnl.gov}
\affiliation{Physics Department, Brookhaven National Laboratory, Upton, NY 11973, USA}

\begin{abstract}

We present an approach for computing the real parts of the nonleptonic $B\to DP$ and $B\to \overline{D}P$ ($P=K,\pi$) decay amplitudes by using lattice QCD methods. While it remains very challenging to calculate the imaginary parts of these matrix elements on the lattice, we stress that their real parts play a significant role in extracting the angle $\gamma$ in the $b{-}d$ unitarity triangle of the CKM matrix. The real part on its own gives a lower bound to the absolute magnitude of the amplitude which is in itself an important constraint for determining $\gamma$. Also the relevant phase can be obtained by using $B$-decays in conjunction with relevant charm decay data. Direct four-point function calculations on the lattice, while computationally demanding, do yield the real part as that is not impeded by the Maiani-Testa theorem. As an approximation, we argue that the chiral expansion of these decays is valid in a framework similar to that of hard-pion chiral perturbation theory. In addition to constructing the leading-order operators, we also discuss the features of the next-to-leading order chiral expansion. These include the contributions from the resonance states, as well as the generic forms of the chiral logarithms.

\end{abstract}

\pacs{11.15.Ha,11.30.Rd,12.38.Gc,12.39.Fe,12.39.Hg}

\maketitle

\section{Motivation and Introduction}\label{sec:intro}

The $B^ +\to D ^0P^+$, $ B^ +\to \Dbar^0 P^+$ decay processes (with $P=K,\pi$) are of significant phenomenological importance. 
These decays can be used for a direct, data-driven, extraction of the $CP$-odd phase $\gamma$ of the $b{-}d$ unitarity triangle 
in the CKM matrix. In principle, given sufficient number of $B$-mesons 
they can provide a determination of $\gamma$ from experiment to an unprecedented accuracy of $\approx$  0.1\% \cite{Asner_rev10}.
In comparison, the projected accuracy of the angle $\beta$ is at best $\approx 0.5\%$, and for $\alpha$ it is likely 
limited to a few percent \cite{RMP09}.  
Using only the charged $B$ meson decays to achieve a precise determination of $\gamma$ is highly valued, since the underlying
decay modes are dominated by tree-level weak-interaction processes.

These  direct methods for deducing $\gamma$ from charged $B$ meson decays involve interference between $D^0$
and $\Dbar^0$ decays to common final states,\footnote{Note that this involves $D^0 - \Dbar^0$ mixing and in the Standard Model, $CP$ violation from this source is assumed to be very small; this is assumed in all $\gamma$  analyses so far \cite{Asner_rev10}.}
 for example
\bea
 	B^{-} \to D^{0} P^{-} \to f P^{-}
	\mbox{ }{\mathrm{and}}\mbox{ }\mbox{ } 
	B^{-} \to \Dbar^{0} P^{-} \to f P^{-}\ ,
\eea
where $f=K_S \pi^0, K^+ K^-, K^- \pi^+, K^- \rho^+, K^{*-} \pi^+$, \emph{etc}.~\cite{Gronau:1990ra,Gronau:1991dp,ADS96,ADS00, GGSZ}.  In particular,
when the final state $f$ is such that decays from $\Dbar^0$ are Cabibbo allowed, but those from $D^0$ are doubly-Cabibbo
suppressed, {\it e.g.} $K^+ \pi^-$,  then in the overall charged $B$ decays, there is a tendency for the interference to be maximal, giving rise to the possibility of large $\cO(1)$ $CP$-asymmetries \cite{ADS96}.  This is important as larger asymmetries
tend to require fewer $B$ mesons for deducing $\gamma$.
 
Since the time these methods were proposed, it has always been recognized and emphasized that studies of charm decays can be very helpful for extracting $\gamma$ \cite{ADS96,Abi98,SS99,ADS00}; in particular, precise knowledge of the branching ratio of the relevant charm decay modes and the strong phase(s) can significantly facilitate the determination of $\gamma$. Specific methods \cite{Atwood:2003mj} have been proposed for studies at charm facilities for this purpose  and great deal of experimental activity has taken place and progress is being made \cite{Rosner:2011sj,Thomas:2010zzj,Malde:2011mk,Ricciardi:2011vk,AP07}.

The methods proposed in Refs.~\cite{ADS96,ADS00} allow
the extraction of $\gamma$, as well as the relevant strong phase difference in $B^{-} \to D^{0} P^{-}$ and $B^{-} \to \Dbar^{0} P^{-}$
amplitudes. In these methods, the branching ratio,
\begin{equation}
\label{eq:br_BtoDK}
	{\rm Br}[B^{-} \to D^{0} P^{-}] ,
\end{equation}
is an essential input. This branching ratio (for $P=K$ or $\pi$) has been experimentally measured with good precision. Due to technical reasons, Br[$B{^-} \to \Dbar^0 P{^-}$] is not accessible to experiment \cite{ADS96}; for this reason
in the method of \cite{ADS96,ADS00} this branching ratio, expressed as the ratio,
\begin{equation}
\label{eq:ratio_r}
	r_{BP} = \frac {{\rm Br}[B{^-} \to \Dbar^0 P{^-}]} {{\rm Br}[B{^-} \to
	{D{^0} P{^-}]}}\ ,
\end{equation}
is treated as an unknown that can be solved for along with $\gamma$. However, determination of this ratio in addition to $\gamma$, places additional demands on the number of $B$ mesons that are needed. For this reason, despite the large statistics of the two $B$-factories [$\sim\cO(10^9)$ $B$ meson samples], $\gamma$ is presently determined to only $\sim\cO(25\%)$. This should be compared with about 3\% for $\beta$, and about 5\% for $\alpha$. To further improve the accuracy on  $\gamma$, inputs from lattice QCD (LQCD) on the ratio in Eq.~(\ref{eq:ratio_r}) would be very useful. In other words if the lattice could provide an accurate value of this ratio, a fewer number of $B$ meson samples will be needed to achieve a given accuracy on $\gamma$.

For the purpose of a lattice study, we define a ``reduced'' ratio which is independent of the CKM matrix elements,
\begin{equation}\label{eq:reduced_r}
	r^{\rm red}_{BP}  
	\equiv \frac{r_{BP}}{V^{\rm combo}_{CKM}}
	= r_{BP}\frac{|V_{cb}^* V_{uq}|^2}{|V_{ub}^* V_{cq}|^2}\ ,
\end{equation}
where $q = s,d$ depending on whether $P=K$ or $\pi$. Needless to say, the study of hadronic weak decays on the lattice continues to represent an outstanding challenge. Exploratory studies \cite{Simone1,Simone2,MartiDKP} initiated in the 80's did not have much success, because of the Maiani-Testa no-go theorem (MTNGT) \cite{Maiani:1990ca}. This theorem states that Euclidean four-point correlators (three sources for external hadrons plus one weak-operator insertion point) always result in the average of in- and out-states, leading to the impossibility of extracting information about the strong phases.  That is, one can only compute the real parts of nonleptonic decay amplitudes from such correlators in a finite volume.\footnote{The $D{-}P$ spectrum in finite volume is rendered discrete,
enabling the extraction of the energy of the excited state which
corresponds to the physical state \cite{Lin:2001ek}.}  For the calculation of $K \to \pi \pi$ on the lattice, one can evade the MTNGT using the  Lellouch-L\"{u}scher (LL) method \cite{Lellouch:2000pv}, and the RBC-UKQCD collaboration is making considerable progress \cite{Qi_ReA0_11,Blum:2011ng} in this avenue. 

On the other hand, the lattice computation of these $B \to D P$ decay amplitudes remains challenging, both because the calculation of the lattice four-point function is computationally demanding to evaluate and because the LL method is only applicable to processes involving elastic final-state scatterings. With the advent of new powerful computers such as the BG/Q, the former difficulty may be overcome in the near future, especially since lattice results for the real part of these amplitudes could provide valuable information on the ratio Eq.~(\ref{eq:reduced_r}), and thereby help in the extraction of $\gamma$ when combined with experimental measurements. For now, we will investigate the use of some approximation methods for tackling these amplitudes.\footnote{It is useful to note that lattice calculation of these $B \to D (\Dbar) P$ amplitudes involve no mixing with lower dimensional operators, ``eye-graphs'' or disconnected diagrams and to that extent are simpler than $K \to 2\pi$ amplitudes in the $\Delta I=1/2$ channel.}

Let us recapitulate that for a determination of the ratio, Eq.~(\ref{eq:reduced_r}), what one needs is the absolute value of the amplitudes for the $B \to D^0 P$ and $B \to \Dbar^0 P$ modes, and not just the real part of the amplitudes that is accessible on the lattice. So what we envision is that the phase of these amplitudes will also  be accessible by combining information from the method of \cite{ADS96,ADS00} with the phase of the relevant charm modes coming from charm studies as briefly alluded to above.

While the strong phase is vital to determining $\gamma$, the real part of the amplitude is still useful in and of itself. This is obtained from direct computation on the lattice from four-point function studies or via approximation methods, directly yielding a lower bound on the absolute value of the amplitude, which would be a valuable constraint on $\gamma$ extraction. Comparing with the progress on $K\to2\pi$ decays \cite{Qi_ReA0_11,Blum:2011ng}, it is reasonable to foresee a lattice calculation which may reach a precision on the real part of this amplitude on the order of 15-20\% within the next five years. With the input of such lattice computations, and the progress in the analysis of the CLEO data for the charm decays \cite{Ricciardi:2011vk}, one can envisage the extraction of $\gamma$ with error around 10\%.

Regarding the approximation methods for lattice studies, we first examine  the possibility of chiral expansion of the real parts of $B^{-}\to D^{0} P^{-} $ and $B^{-}\to \Dbar^{0} P^{-}$ amplitudes, in the framework of heavy-meson chiral perturbation theory (HM\chpt)  which merges heavy quark effective theory and chiral perturbation theory (\chpt) \cite{Wise:1992hn,Burdman:1992gh,Yan:1992gz,Cho:1992gg,Cho:1992cf,Manohar:2000dt}. The presence of the $b$ and $c$ quarks, both heavier than $\Lambda_{\rm QCD}$, allows for systematic expansion in terms of $\Lambda_{\rm QCD}/m_{b,c}$ ($m_{b,c}$ is the $b,c$ quark mass). This expansion has already been used both in lattice determinations to leptonic and semi-leptonic decays, as well as in the continuum. Combined with the chiral expansion, it leads to a powerful tool for extrapolating lattice data to the physical pion mass. This extrapolation will still be an essential step in lattice calculations in the foreseeable future, since most lattice simulations are not yet performed at the physical pion mass.

To begin, we examine the validity of HM\chpt\ for the processes we are interested in. In the limit where both the $b$ and $c$ quarks are treated as static, resulting in a soft final-state Goldstone boson, this approach is valid. However, this limit is far from the physical regime, and such an extrapolation would introduce significant systematic errors. Therefore, the straightforward applicability of \chpt\ is questionable ({\it i.e.} it would be a poor approximation with rather large errors). On the other hand, if we perform simulations near the physical kinematic point of the decays of interest [$B\to D K(\pi)$], the emerging $D$ meson and the Goldstone boson are hard, with $p\sim 2$ GeV.\footnote{Note that this implies large discretization errors of the form $(ap)^n$, with $n>0$, and as such it would require either very fine lattices or a choice of action which would largely suppress these errors.} 

The appearance of hard external momenta does not, as one may initially assume, lead to a breakdown in the chiral expansion. It was recently shown that treating the $s$-quark as heavy and using SU(2) \chpt\ works quite well for chiral extrapolations \cite{Antonio07,Allton08}. This method can be generalised to processes in which external pions have hard momenta, and applications have appeared in analyzing $K_{\ell3}$ decays \cite{Flynn:2008tg}, $K\to  2\pi$ \cite{Bijnens:2009yr}, as well as extensions to semi-leptonic $B$-decays \cite{Bijnens:2010ws}. A central concern in all of these applications is how well \chpt\ works in the presence of hard momenta. In particular, there is evidence that the hard pion does not spoil the chiral logarithms, at least to next-to-leading order (NLO) \cite{ColangeloLAT11}\footnote{As noted in Ref.~\cite{Bijnens:2009yr}, hard-pion $\chi$PT may not be applicable to the extraction of the imaginary parts of the $K\to \pi\pi$ amplitudes.  We will comment on this issue for nonleptonic $B$ decays in Sec.~\ref{sec:hardPi}.}.

Applying this to processes involving $D$ mesons is straightforward, and a key result is that the hard momenta of the external mesons (both the $D$ and the pion) will be absorbed into a redefinition of the low-energy constants (LEC's), and thus all remaining quantities will be soft. Thus we can still treat the $D$ meson using the non-relativistic approach of HM\chpt, so that corrections arising in the $D$ sector will arise at $\cO(\Lambda_{\rm QCD}/M_D)$, as usual.

To investigate the relevant $B$ decay processes, we are interested in the following current-current, $\Delta b=1$, operators ($\alpha,\beta$ are  color indices)
\bea
	Q^{b\to c,i}_1 &=&
	 (\qbar^i_\alpha \gamma^\mu(1-\gamma_5)b_\alpha)
	(\cbar_\beta\gamma_\mu(1-\gamma_5) u_\beta)\ ,\label{eq:btoc1}\\
	Q^{b\to c,i}_2 &=&
	 (\qbar^i_\alpha\gamma^\mu(1-\gamma_5) b_\beta)
	(\cbar_\beta \gamma_\mu(1-\gamma_5)u_\alpha)\ ,\label{eq:btoc2}\\
	Q^{b\to \cbar,i}_1 &=&
	(\qbar^i_\alpha \gamma^\mu(1-\gamma_5)b_\alpha)
	(\ubar_\beta \gamma_\mu(1-\gamma_5)c_\beta)\ ,
	\label{eq:btocbar1}\\
	Q^{b\to \cbar,i}_2 &=&
	(\qbar^i_\alpha\gamma^\mu(1-\gamma_5) b_\beta)
	(\ubar_\beta \gamma_\mu(1-\gamma_5)c_\alpha)\ .\label{eq:btocbar2}
\eea
For the decay channels $B^{-}\to D^{0}K^{-}$ or $B^{-}\to\Dbar^{0} K^{-}$, we will set $q^i = s$ and for 
$B^{-}\to D^{0}\pi^{-}$ or $B^{-}\to\Dbar^{0}\pi^{-}$ we have $q^i = d$. The corresponding effective Hamiltonian for these decays is
\be\label{eq:Heff}
	H_{\rm eff} = 
	\frac{G_F}{\sqrt{2}} \sum_{j=1,2}\sum_{i=d,s} [
	V_{cb}^*V_{uq^i} C_j(\mu) Q^{b\to c,i}_j 
	+ V_{ub}^*V_{cq^i} C_j(\mu) Q^{b\to \cbar,i}_j
	+ {\rm h.c.}]\ .
\ee	

We will focus on the nonleptonic decays which have the underlying processes $b \to c\ubar d,\ b \to c\ubar s,\ b \to u\cbar d,$ and $b \to u\cbar s$. The first two will be mapped onto different operators in the chiral theory than the final two, because they belong to different irreducible representations under the chiral transformation.  Furthermore, a chiral field which creates a heavy-light meson with a $c$ quark is \emph{not} a field which destroys a heavy-light meson with a $\cbar$ anti-quark. We will discuss the details of these operators in the chiral effective theory in Sec.~\ref{sec:ChPT_BtoDK}. 

The outline for this paper is as follows. First, in Sec.~\ref{sec:hl-chpt} we present an introduction to HM\chpt\ for $B,D$ and $\Bbar,\Dbar$ mesons. In Sec.~\ref{sec:ChPT_BtoDK}, we construct the \chpt\ operators for the quark-level operators in Eqs.~(\ref{eq:btoc1})-(\ref{eq:btocbar2}). We then treat the leading-order calculation of $B\to DP$ and $B\to \Dbar P$ and relate them to the unphysical $B\to D$ and $B\to \Dbar$ processes. In Sec.~\ref{sec:resonance}, we discuss the tree-level resonance (initially either a $B^*$ or $D^*$) contributions to the nonleptonic $B$ decays in the framework of HM\chpt. Finally we sketch the steps required for a full one-loop calculation in Sec.~\ref{sec:hardPi} and conclude in Sec.~\ref{sec:disc}.

\section{Heavy-Meson Chiral Perturbation Theory}\label{sec:hl-chpt}
The strong-interaction chiral Lagrangian for the Goldstone bosons is (the $\eta^{\prime}$ is already integrated out)~\cite{Gasser:1983yg,Gasser:1984gg}
\beq
\label{eq:goldstone_chiral_lagrangian}
 {\mathcal{L}}_{\mathrm{G}} =
 \frac{f^{2}}{8} {\mathrm{Tr}} \left (
 \partial_{\mu} \Sigma \partial^{\mu} \Sigma^{\dagger} \right )
 + \frac{1}{4}\mu f^{2} {\mathrm{Tr}} 
   \left ( \ampl \Sigma + \ampl \Sigma^{\dagger} \right ) ,
\eeq
where $\mu$ is a low-energy constant (LEC) related to the chiral condensate, $\ampl$ is the light-quark mass matrix,
\beq
\label{eq:mass_matrix}
 \ampl = {\mathrm{diag}} 
  \left ( m_{u}, m_{d}, m_{s} \right ) ,
\eeq
$\Sigma={\mathrm{exp}}(2 i \Phi/f)$ is the non-linear Goldstone particle field, with $\Phi$ being the matrix containing the standard Goldstone fields, and we use a normalization for $f$ such that $f_\pi\approx130.7$~MeV. Under an ${\mathrm{SU}}(3)_{{\mathrm{L}}}\otimes {\mathrm{SU}}(3)_{{\mathrm{R}}}$ chiral rotation, $\Sigma$ transforms as
\beq\label{eq:Sigma_transform_rule}
	\Sigma \longrightarrow L \: \Sigma \: R^{\dagger},
	\mbox{ }{\mathrm{where}}\mbox{ } 
	L \in {\mathrm{SU}}(3)_{{\mathrm	{L}}},
	\mbox{ }{\mathrm{and}}\mbox{ } R \in 
	{\mathrm{SU}}(3)_{{\mathrm{R}}}.
\eeq

To account for the light-quark dynamics in heavy-light mesons, one can combine the formulations for heavy quark effective theory (HQET) and \chpt\ into HM\chpt~\cite{Wise:1992hn,Burdman:1992gh,Yan:1992gz,Cho:1992gg,Cho:1992cf,Manohar:2000dt}. There is a $U(2m)$ spin-flavor symmetry on the heavy-quark side for $m$ heavy quarks, and the standard (broken) ${\mathrm{SU}}(3)_{{\mathrm{L}}}\otimes {\mathrm{SU}}(3)_{{\mathrm{R}}}$ chiral symmetry for the light quarks. 

We will sketch the relevant details for constructing HM\chpt, using the notation of Ref.~\cite{Aubin:2005aq}. First, we have the field which destroys (creates) a heavy-light meson
\be
	H_{v,a}^{(Q)} = 
	\left ( \frac{1+\vslash}{2} \right ) \left(
	 \gamma^\mu \cV^{*(Q)}_{\mu,a} -
	\gamma_5 \cP^{(Q)}_a \right ) \ ,\quad
	\Hbar_{v,a}^{(Q)} \equiv 
	\gamma^{0} H_{a}^{(Q)\dagger} \gamma^{0} = 
	\left ( \gamma^\mu \cV^{*(Q)\dagger}_{\mu,a} +  
	\gamma_5 \cP^{(Q)\dagger}_a \right ) \left ( \frac{1+\vslash}{2} \right ) \ ,
\ee
where $a$ is the light quark flavor index, $Q$ is the heavy-quark index, and $v$ is the four-velocity of the heavy-light meson. We use $\cP$ for the heavy-light pseudoscalar field and $\cV^*$ for the heavy-light vector field. For the heavy-light fields with heavy anti-quarks, we have \cite{Grinstein:1992qt}
\be
	H^{(\Qbar)}_{v,a} 
	= \left(\gamma^\mu \cV_{\mu,a}^{*(\Qbar)} - \gamma_5 \cP^{(\Qbar)}_{a}
	\right)\left(\frac{1-\vslash}{2}\right)\, \quad
	\Hbar^{(\Qbar)}_{v,a} \equiv \gamma^{0} H_{a}^{(\overline{Q})\dagger} \gamma^{0}
	= \left(\frac{1-\vslash}{2}\right)
	\left(\gamma^\mu \cV_{\mu,a}^{*(\Qbar)\dag} + \gamma_5 \cP^{(\Qbar)\dag}_{a}
	\right)\ .
\ee
It is convenient when dealing with both charm and bottom quarks and antiquarks to combine them into multiplets which transform under the $U(4)$ spin/flavor symmetry, 
\be
	H_{Q,v,a} = \left(\begin{matrix}
	H^{(b)}_{v,a} \\ H^{(c)}_{v,a}
	\end{matrix}\right)\ , \qquad
	H_{\Qbar,v,a} = \left(\begin{matrix}
	H^{(\bbar)}_{v,a} \\ H^{(\cbar)}_{v,a}
	\end{matrix}\right)\ .
\ee

Suppressing the light flavour and velocity indices, under the heavy-quark spin/flavour transformation $S \in U(4)$, and the unbroken light-flavour transformation $\mathbb{U}(x)$, the above fields transform as 
\bea
  && H_{Q}(x) \to  S \: H_{Q}(x) \: \mathbb{U}^{\dagger}(x)\ , \mbox{ }
  \overline{H}_{Q}(x) \to  \mathbb{U}(x) \: \overline{H}_{Q}(x) \: S^{\dagger}\ , \nonumber\\
&&\nonumber\\
\label{eq:HL_field_transformation}
  && H_{\overline{Q}}(x) \to  \mathbb{U}(x) \: H_{\overline{Q}}(x) \: S^{\dagger} \ , \mbox{ }
  \overline{H}_{\overline{Q}}(x) \to  S \: \overline{H}_{\overline{Q}}(x) \: \mathbb{U}^{\dagger}(x)\ , 
\eea
where $\mathbb{U}(x)$ is a function of $L$, $R$ and $\Phi(x)$.

The Goldstone bosons couple to the heavy-light mesons in HM\chpt\ via the non-linear realisation
\beq
 \sigma = \sqrt{\Sigma} = {\mathrm{e}}^{i \Phi/f} ,
\eeq
which transforms as 
\beq
  \label{eq:Udef}
  \sigma(x) \to  L \: \sigma(x) \:  \mathbb{U}^{\dagger}(x) = \mathbb{U}(x) \: \sigma(x) \: R^{\dagger}\,, \, \,
  \sigma^{\dagger}(x) \to R \: \sigma^{\dagger}(x) \: \mathbb{U}^{\dagger}(x) = \mathbb{U}(x) \: \sigma^{\dagger}(x) \: L^{\dagger}\,.
\eeq
Due to the properties of the heavy-light meson fields in Eq.~(\ref{eq:HL_field_transformation}), 
it is convenient to define objects involving the $\sigma$ field that transform 
only with $\mathbb{U}$ and $\mathbb{U}^\dagger$. The two 
possibilities with a single derivative are
\begin{eqnarray}
  \mathbb{V}_{\mu} & = & \frac{i}{2} \left[ \sigma^{\dagger} \partial_\mu
   \sigma + \sigma \partial_\mu \sigma^{\dagger}   \right] \ ,
 \\
&& \nonumber\\
  \mathbb{A}_{\mu} & = & \frac{i}{2} \left[ \sigma^{\dagger} \partial_\mu
   \sigma - \sigma \partial_\mu \sigma^{\dagger}   \right] \ .
\end{eqnarray}
The Lorentz vector $\mathbb{V}_{\mu}$ can be
combined with the derivative to form a 
covariant derivative acting on the heavy-light field or its conjugate: 
\begin{eqnarray}\label{eq:Ddef}
	H_{Q,v,b} \leftvec D^{ba}_\mu  
	&\equiv& \partial_\mu H_{Q,v,a} + i H_{Q,v,b} \mathbb{V}_{\mu}^{ba}\ , 
	\nonumber \\
&& \nonumber\\ 
	\rightvec D^{ab}_\mu \overline{H}_{Q,v,b} 
	&\equiv& \partial_\mu \overline{H}_{Q,v,a} - 
	i \mathbb{V}_{\mu}^{ab} \overline {H}_{Q,v,b}\ ,
\end{eqnarray}
with implicit sums over repeated indices, and similarly for the $H_{\Qbar,a}$ fields.
The covariant derivatives and $\mathbb{A}_\mu$
transform under the unbroken light-flavour symmetry as
\begin{eqnarray}\label{eq:Dtransf}
	H \leftvec D_\mu \to  (H \leftvec D_\mu )\mathbb{U}^\dagger\ , 
	&&
	\rightvec D_\mu \overline{H} \to  \mathbb{U} (\rightvec D_\mu \overline{H})\ , \\
&&\nonumber\\
	 \mathbb{A}_\mu &\to&  \mathbb{U} \mathbb{A}_\mu \mathbb{U}^\dagger\ ,
\end{eqnarray}
where we have dropped all the indices for simplicity.

The leading-order chiral Lagrangian is given by $\cL_{\rm LO} = \cL_{\rm G} + \cL_{\rm HL,1}$, where
\begin{equation}\label{eq:L1}
  \cL_{\rm HL,1}  =  -i  \Tr(\Hbar H v\negcdot \leftvec D )
  + g_\pi \Tr(\Hbar H \gamma^{\mu}\gamma_5 \mathbb{A}_{\mu}) \ .
\end{equation}
$\Tr$ means the complete trace over light quark flavor indices, heavy quark flavor indices, and, where relevant, Dirac indices. Since $\Hbar$ and $H$ always appear together in the Lagrangian, we treat $\Hbar H$ as a matrix in light-quark flavor space: $(\Hbar H)_{ab} \equiv \Hbar_a H_b$.  The axial coupling $g_{\pi}$ in the above Lagrangian determines the $B^{\ast}{-}B{-}$Goldstone and $D^{\ast}{-}D{-}$Goldstone interaction strength. Its value, $g_{\pi}\approx 0.45$, has recently been computed using unquenched lattice QCD~\cite{Ohki:2008py,Becirevic:2009yb,Detmold:2011bp}.

At the next-to-leading order (NLO), the Lagrangian contains a number of additional terms~\cite{Boyd:1995pa,Stewart:1998ke,Aubin:2005aq}.  Among these terms, only one of them,
\beq
\label{eq:HQS_breaking_term}
  \lambda_{2} \Tr \left ( \frac{1}{M_{\calP}}\overline{H} \sigma_{\mu\nu} H \sigma^{\mu\nu} \right ), 
\eeq
is relevant to this paper ($\lambda_{2}$ is a LEC). This operator breaks the heavy-quark spin symmetry and results in the $B^{\ast}{-}B$ and $D^{\ast}{-}D$ mass splittings.  Notice that $M_{\calP}$ is taken to be the corresponding $B$ and $D$ meson masses in this work, and we do not include other effects related to the breaking of heavy-quark flavour symmetry.

\section{The chiral expansion for $B \to D K (\pi)$ amplitudes}\label{sec:ChPT_BtoDK}
The difficulty in the use of \chpt\ in computations for $B \to D P$ decay amplitudes originates in the large momenta carried by the final state hadrons. In general, the chiral expansion is known to be applicable only to processes involving momenta well below the chiral symmetry breaking scale.  On the other hand, it has been established recently that \chpt\ can be valid for amplitudes containing hard final state particles~\cite{Flynn:2008tg,Bijnens:2009yr,Bijnens:2010ws,Bijnens:2010jg}. One important point in such procedures is that the LEC's are no longer universal quantities for a fixed number of sea quarks.  Rather, they depend on the hard momentum scale which results from either the kinematics or the mass of the external particles. 

This procedure of separating the hard scales in a process is described in detail in the references given above. The key point in this separation lies with the derivative couplings that give rise to momentum dependence in \chpt\ calculations. When these momenta are external and hard, they can be absorbed into the LEC's of the theory. We will discuss this procedure explicitly with an example diagram for the process $B\to DP$ in Sec.~\ref{sec:hardPi}.

First we discuss the construction of the \chpt\ weak operators corresponding to those in 
Eqs.~(\ref{eq:btoc1})--(\ref{eq:btocbar2}).
Omitting the colour indices which do not play a role in \chpt, these operators can be written as
\bea
     && Q^{b\to c,i}
	= \left ( \qbar^{i}_{L} \mbox{ }\Gamma_{1}\mbox{ } b \right ) \left ( \cbar \mbox{ }\Gamma_{2}\mbox{ } u_{L} \right ) , \nonumber\\
&&\nonumber\\
\label{eq:Q_chiral_form}
     && Q^{b\to \overline{c},i}
	= \left ( \qbar^{i}_{L} \mbox{ } \overline{\Gamma}_{1} \mbox{ } b \right ) \left ( \ubar_{L} \mbox{ } \overline{\Gamma}_{2}\mbox{ } c \right ) ,
\eea
where $q^{i} = d$ or $s$, and 
\bea
  && q_{L} = \left ( \frac{1 - \gamma_{5}}{2} \right )\mbox{ } q , \nonumber\\
&& \nonumber\\
  && \Gamma_{1} = \Gamma_{2}  =  \overline{\Gamma}_{1} = \overline{\Gamma}_{2} = \gamma_{\mu} (1 - \gamma_{5}) .
\eea
Under the ${\mathrm{SU}}(3)_{{\mathrm{L}}} \otimes {\mathrm{SU}}(3)_{{\mathrm{R}}}$
chiral symmetry group, $Q^{b\to c,i}$ is in the $({\bf 8_L,1_R})$ representation, while $Q^{b\to \overline{c},i}$ is in the 
$({\bf \bar{6}_L,1_R})$ representation.  To bosonise these operators, we promote $\Gamma_{1,2}$ and $\overline{\Gamma}_{1,2}$ to
be spurion fields which transform as
\bea
 && \Gamma_{1} \to L\mbox{ }\Gamma_{1}\mbox{ }S^{\dagger} , \mbox{ }\mbox{ }
    \Gamma_{2} \to S\mbox{ }\Gamma_{2}\mbox{ }L^{\dagger} , \nonumber\\
&& \nonumber\\
\label{eq:spurion_transformation}
  && \bar{\Gamma}_{1} \to L\mbox{ }\overline{\Gamma}_{1}\mbox{ }S^{\dagger} , \mbox{ }\mbox{ }
     \bar{\Gamma}_{2} \to L\mbox{ }\overline{\Gamma}_{2}\mbox{ }S^{\dagger} ,
\eea
under the heavy-quark spin/flavour and chiral rotations. This renders the operators in Eq.~(\ref{eq:Q_chiral_form}) invariant with respect to such transformations. We then find the bosonisation results in the leading order (LO) operators
\beq
\label{eq:chiral_O_original}
 \op_{\chi,i} = \sum_{x} \left \{ \alpha_{1,x} {\mathrm{Tr}}_{{\mathrm{D}}}
      \left [ \left ( \sigma_{1k} \overline{H}^{(c)}_{v^{\prime},k} \right ) \Gamma_{2} 
       \Xi^{\prime}_{x} \Xi_{x} \Gamma_{1} \left ( H^{(b)}_{v,l} \sigma^{\dagger}_{li} 
            \right ) \right ] + \alpha_{2,x} {\mathrm{Tr}}_{{\mathrm{D}}}
      \left [ \left ( \sigma_{1k} \overline{H}^{(c)}_{v^{\prime},k} \right ) \Gamma_{2} 
       \Xi^{\prime}_{x} \right ] 
       {\mathrm{Tr}}_{{\mathrm{D}}}
      \left [  \Xi_{x} \Gamma_{1} \left ( H^{(b)}_{v,l} \sigma^{\dagger}_{li} 
            \right ) \right ] \right \} ,
\eeq
for $Q^{b\to c,i}$, and
\beq
\label{eq:chiral_Obar_original}
 \overline{\op}_{\chi,i} = \sum_{x} \left \{ \overline{\alpha}_{1,x} {\mathrm{Tr}}_{{\mathrm{D}}}
      \left [ \Xi^{\prime}_{x} \Gamma_{2} \left ( \overline{H}^{(\overline{c})}_{v^{\prime},k} \sigma_{k1}^{\dagger}  \right ) 
          \Xi_{x} \Gamma_{1} \left ( H^{(b)}_{v,l} \sigma^{\dagger}_{li} 
            \right ) \right ] + \overline{\alpha}_{2,x} {\mathrm{Tr}}_{{\mathrm{D}}}
      \left [ \Xi^{\prime}_{x} \Gamma_{2} \left ( \overline{H}^{(c)}_{v^{\prime},k} \sigma_{k1}^{\dagger} \right ) \right ] 
       {\mathrm{Tr}}_{{\mathrm{D}}}
      \left [  \Xi_{x} \Gamma_{1} \left ( H^{(b)}_{v,l} \sigma^{\dagger}_{li} 
            \right ) \right ] \right \} ,
\eeq
for $Q^{b\to \overline{c},i}$,
where ${\mathrm{Tr}}_{{\mathrm{D}}}$ means the trace in Dirac space, and the summation over repeated indices are assumed. 
The symbols $\Xi^{\prime}_{x}$ and $\Xi_{x}$ are all possible pairs of Dirac structures
allowed by symmetries~\cite{Detmold:2006gh},
\bea
 \left \{ \Xi^{\prime}_{x},\Xi_{x} \right \} &=&
  \bigg \{ 
       \left \{ 1, 1 \right \}, 
       \left \{ \gamma_{\nu}, \gamma^{\mu} \right \}, 
       \left \{ \slash\!\!\! v^{\prime}, \slash\!\!\! v \right \}, 
       \left \{ \slash\!\!\! v^{\prime}, 1\right \}, 
       \left \{ 1, \slash\!\!\! v\right \}, 
       \left \{ \sigma_{\mu\nu}, \sigma^{\mu\nu}\right \},\nonumber\\
\label{eq:Dirac_structures}
    && \mbox{ }\mbox{ }\mbox{ }
       \left \{ \gamma_{5}, \gamma_{5} \right \}, 
       \left \{ \gamma_{\mu}\gamma_{5}, \gamma^{\mu}\gamma_{5}\right \}, 
       \left \{ \slash\!\!\! v^{\prime} \gamma_{5}, \slash\!\!\! v \gamma_{5}\right \}, 
       \left \{ \slash\!\!\! v^{\prime} \gamma_{5}, \gamma_{5}\right \}, 
       \left \{ \gamma_{5}, \slash\!\!\! v \gamma_{5}\right \} \bigg \}.
\eea
In particular, the positions of these Dirac structures in HM\chpt\ weak operators are constrained by heavy-quark 
spin/flavour symmetry.  They have to be inserted to account for light-quark and gluon dynamics.

\begin{figure}[t]
\begin{center}
\includegraphics[width=5in]{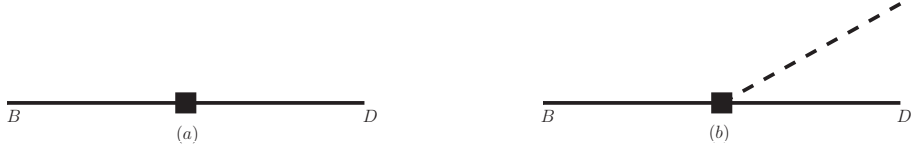}
\caption{Tree-level diagrams contributing to (a) $B\to D$ and (b) $B\to DP$ at lowest order, with no insertions of the strong Lagrangian. The box is the weak operator, the solid line is a heavy-light pseudoscalar (either $B$ or $D$), and the dashed line is the light meson $P$.}
\label{fig:BtoDtreeOrderOne}
\end{center}
\end{figure}

Performing the Dirac traces in Eqs.~(\ref{eq:chiral_O_original}) and (\ref{eq:chiral_Obar_original}), we obtain
\bea
 \op_{\chi,i} = &&
  \left [ \beta_{1} + 
   \left ( \beta_{1} + \beta_{2}\right ) (v^{\prime}\cdot v)
  \right ] 
  \left [ \left ( \sigma_{1k}  \calP^{(c)\dagger}_{k}\right ) 
          \left ( \calP^{(b)}_{l} \sigma^{\dagger}_{li} \right )
  \right ]\nonumber\\
  &+&
  \left [ \left ( \beta_{1} - \beta_{2}\right ) v^{\prime\mu}
          - \beta_{1} v^{\mu}
  \right ] 
  \left [ \left ( \sigma_{1k}  \calP^{(c)\dagger}_{k}\right ) 
          \left ( \calV^{\ast(b)}_{\mu,l} \sigma^{\dagger}_{li} \right )
  \right ]\nonumber\\
  &+&
  \left [ \beta_{1} v^{\prime\mu}
          - \left ( \beta_{1} + \beta_{2}\right ) v^{\mu}
  \right ] 
  \left [ \left ( \sigma_{1k}  \calV^{\ast(c)\dagger}_{\mu,k}\right ) 
          \left ( \calP^{(b)}_{l} \sigma^{\dagger}_{li} \right )
  \right ]\nonumber\\
  &-& 4
  \left [ \left ( \beta_{1} - \beta_{2}\right )
         + \beta_{1} (v^{\prime}\cdot v)
  \right ] 
  \left [ \left ( \sigma_{1k}  \calV^{\ast(c)\dagger}_{\mu,k}\right ) 
          \left ( \calV^{\ast(b)\mu}_{l} \sigma^{\dagger}_{li} \right )
  \right ], \nonumber\\
&&\nonumber\\
 \overline{\op}_{\chi,i} = &&
  \left [ \overline{\beta}_{1} + \overline{\beta}_{2} (v^{\prime}\cdot v)
  \right ] 
  \left [ \left ( \calP^{(\bar c)\dagger}_{k} \sigma_{k1}^{\dagger}  \right ) 
          \left ( \calP^{(b)}_{l} \sigma^{\dagger}_{li} \right )
  \right ]\nonumber\\
  &-&
  \left [ \overline{\beta}_{2} v^{\prime\mu} - \left ( \overline{\beta}_{1} + \overline{\beta}_{5} \right ) v^{\mu}
       - \overline{\beta}_{3} (v^{\prime}\cdot v) v^{\mu}
  \right ] 
  \left [ \left (\calP^{(\bar c)\dagger}_{k}  \sigma_{k1}^{\dagger} \right ) 
          \left ( \calV^{\ast(b)}_{\mu,l} \sigma^{\dagger}_{li} \right )
  \right ]\nonumber\\
  &+&
  \left [ \overline{\beta}_{1} v^{\prime\mu} - \overline{\beta}_{2} v^{\mu}
  \right ] 
  \left [ \left (  \calV^{\ast(\bar c)\dagger}_{\mu,k} \sigma_{k1}^{\dagger} \right ) 
          \left ( \calP^{(b)}_{l} \sigma^{\dagger}_{li} \right )
  \right ]\nonumber\\
\label{eq:chiral_O_expanded}
  &+& 
  \left [ 4 \overline{\beta}_{2} - \overline{\beta}_{3}
     - 2 \left ( \overline{\beta}_{1} + \overline{\beta}_{4} +\overline{\beta}_{5} \right ) 
      (v^{\prime}\cdot v)
  \right ] 
  \left [ \left (  \calV^{\ast(\bar c)\dagger}_{\mu,k} \sigma_{k1}^{\dagger} \right ) 
          \left ( \calV^{\ast(b)\mu}_{l} \sigma^{\dagger}_{li} \right )
  \right ], 
\eea
where $\beta_{i}$ is a linear combination of $\alpha_{1,x}$ and $\alpha_{2,x}$ while
$\overline{\beta}_{i}$ is a linear combination of $\overline{\alpha}_{1,x}$ and $\overline{\alpha}_{2,x}$.  At the lowest
order in the chiral expansion, only the first terms in the above operators contribute to $B\to D P$
and $B \to \overline{D} P$ processes.  It is straightforward to demonstrate that if we evaluate the diagrams in Fig.~\ref{fig:BtoDtreeOrderOne} at leading order,
\bea
 && \la D^{0} K^{-} | \op_{\chi,s} | B^{-}\ra = \la D^{0} \pi^{-} | \op_{\chi,d} | B^{-}\ra
      = \frac{i}{f} \la D^{-} | \op_{\chi,s} | B^{-}\ra ,\nonumber\\
&& \nonumber\\
\label{eq:LO_chiral_expansion}
 && \la \overline{D}^{0} K^{-} | \overline{\op}_{\chi,s} | B^{-}\ra = \la \overline{D}^{0} \pi^{-} | \overline{\op}_{\chi,d} | B^{-}\ra
      = \frac{i}{f} \la D^{-} | \overline{\op}_{\chi,s} | B^{-}\ra 
      \ .
\eea
From Eq.~(\ref{eq:chiral_O_expanded}), it is clear that beyond the
LO, the chiral expansion may become very different for $B^{-}\to D^{0} P^{-}$ and $B^{-}\to \overline{D}^{0} P^{-}$ amplitudes. In the 
next two sections, we will discuss the generic features of these amplitudes at the NLO and leave the details to a future publication.

\section{Resonance Contributions}\label{sec:resonance}

\begin{figure}[t]
\begin{center}
\includegraphics[width=5in]{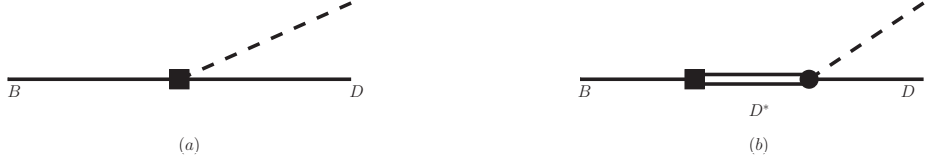}
\caption{Tree-level correlators contributing to $B\to DP$. The box is the weak operator, with (a) being the direct $B\to DP$ term and (b) being the term with an intermediate resonance (here a $D^*$).}
\label{fig:BtoDtreeOrderPpi}
\end{center}
\end{figure}

In this section we discuss one generic feature of $B\to D P$ 
correlators and amplitudes, namely, the resonance contribution.\footnote{The conclusion
presented in this section is also valid for $B\to \overline{D} P$ decays.}  This is partly
incorporated in HM\chpt\ via the inclusion of the vector heavy-light
mesons.  Figure~\ref{fig:BtoDtreeOrderPpi}(b) shows a typical
diagram in which a resonance ($D^{\ast}$ in this case) appears
in the $B\to D P$ correlators.
One can also include heavier resonances in the effective
theory~\cite{Becirevic:2007dg}, but it is beyond the scope of this
paper.  Here we will address the issue regarding the contribution from
the resonance in the time-momentum representation of
correlators.  To avoid complications arising from the formulation of
HQET and HM\chpt\ in Euclidean space~\cite{Aglietti:1992in},
we work in Minkowski space with the comment that we also carried out a
similar calculations by modelling the heavy-light
mesons as relativistic particles in Eulidean space and obtained the
same conclusions presented in this section.

We first set up the calculation for the LO correlator in
Fig.~\ref{fig:BtoDtreeOrderPpi}(a).  To mimic the setting for most
lattice calculations, we integrate over
the spatial volume for the positions of external $B$, $D$ and pion
(kaon) sources/sinks, {\it i.e.}, we perform a Fourier transform
for the spatial directions for each of the external points.   On the other hand,
we fix the location of the weak operator (the square in the
diagram) to be at the origin.  To be consistent with the notation in Sec.~\ref{sec:ChPT_BtoDK},
we denote the velocity of $B$ and $B^{\ast}$ by $v$ and that of 
$D$ and $D^{\ast}$ by $v^{\prime}$.  For simplicity, the velocity $v$
is chosen to be 
\beq
\label{eq:B_velocity}
 v = \left ( 1, \vec{0}\right ) ,
\eeq
and the time-ordering is implemented as
\beq
\label{eq:time-ordering}
 t_{B} < 0 < t_{D} \le t_{P}, 
\eeq
where $t_{B,D,P}$ is the temporal locations of the $B,D,P$ mesons, respectively. 
Using the Feynman rules derived from the HM\chpt\ Lagrangian and the weak operators 
in Eqs.~(\ref{eq:L1}) and (\ref{eq:chiral_O_expanded}), the result for the contribution
from this diagram in the correlator is 
\bea
\label{eq:LO_correlator}
 C_{{\mathrm{LO}}} &=& \frac{g_{BDP}}{f} \left ( \frac{1}{2} \theta(-t_{B}) \right ) 
 \left ( \frac{1}{2 v^{\prime}_{0}} \theta(t_{D}) \exponential^{-i \overline{\delta}_{D} t_{D} } \right )  
 \left ( 
   \frac{\exponential^{-i \omega_{P} t_{P}}}{2 \omega_{P}} \right ) \nonumber \\
&& \nonumber \\
 &=& \frac{g_{BDP}}{f} \left ( \frac{1}{2} \right ) 
 \left ( \frac{\exponential^{-i \overline{\delta}_{D} t_{D} } }{2 v^{\prime}_{0}}\right )  
 \left ( 
   \frac{\exponential^{-i \omega_{P} t_{P}}}{2 \omega_{P}} \right ) ,
\eea
where
\beq
\label{eq:mass_def1}
  \overline{\delta}_{D} = \vec{v^{\prime}} \cdot \vec{p}_{D},
 \mbox{ }\mbox{ }{\mathrm{and}}\mbox{ }\mbox{ } 
 \omega_{P} = \sqrt{M^{2}_{P} + \vec{p}^{2}_{P}} ,
\eeq
with $\vec{p}_{D}$ and $\vec{p}_{P}$ denoting the spatial momenta of the $D$ and the Goldstone boson.
The coupling $g_{BDP}$ is one of the linear combinations of the LEC's $\beta_{i}$ in 
Eq.~(\ref{eq:chiral_O_expanded}). 

\begin{figure}[t]
\begin{center}
\includegraphics[width=3in]{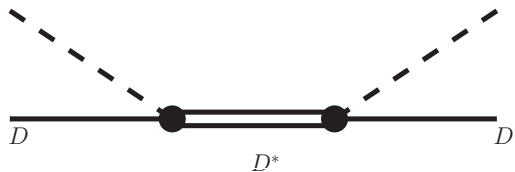}
\caption{A contribution to the process $DP\to DP$ involving a single resonance.}
\label{fig:DPtoDP}
\end{center}
\end{figure}

Next, we discuss the correlator depicted in Fig.~\ref{fig:BtoDtreeOrderPpi}(b).  This diagram is
calculated by integrating over the entire space-time for the location of the strong vertex (denoted
by the circle).  It leads to the result
\bea
\label{eq:resonance_correlator}
 C_{{\mathrm{res}}} &=& \frac{ g_{BD^{\ast}} \left ( i g_{\pi}\right )}{f} \left ( \frac{1}{2} \theta(-t_{B}) \right ) 
 \left ( \frac{1}{2 v^{\prime}_{0}} \theta(t_{D}) \exponential^{-i \overline{\delta}_{D} t_{D} } \right )  
 \left ( 
   \frac{\exponential^{-i \omega_{P} t_{P}}}{2 \omega_{P}} \right )
 \left [
  \frac{\exponential^{i (\omega_{P} + \overline{\delta}_{D} - \overline{\Delta}_{DP} ) t_{D}} - 1}
   {2 i  v^{\prime}_{0} (\omega_{P} + \overline{\delta}_{D} - \overline{\Delta}_{DP} )}
 \right ] \nonumber \\
&& \nonumber \\
 &=& \frac{g_{BD^{\ast}} g_{\pi} }{f}\left ( \frac{1}{2} \right ) 
 \left ( \frac{\exponential^{-i \overline{\delta}_{D} t_{D} }}{2 v^{\prime}_{0}}  \right )  
 \left ( 
   \frac{\exponential^{-i \omega_{P} t_{P}}}{2 \omega_{P}} \right )
 \left [
  \frac{\exponential^{i (\omega_{P} + \overline{\delta}_{D} - \overline{\Delta}_{DP} ) t_{D}} - 1}
   { 2  v^{\prime}_{0} (\omega_{P} + \overline{\delta}_{D} - \overline{\Delta}_{DP} )}
 \right ] ,
\eea
where 
\beq
 \overline{\Delta}_{DP} = \vec{v}^{\prime} \cdot \left ( \vec{p}_{D} + \vec{p}_{\pi} \right )
     + \frac{\Delta_{D}}{v^{\prime}_{0}} ,
\eeq
with $\Delta_{D}$ denoting the $D^{\ast}{-}D$ mass splitting resulting from the heavy-quark
spin symmetry breaking term in Eq.~(\ref{eq:HQS_breaking_term}).  When the final-state momenta
are tuned such that
\beq
 \omega_{P} + \overline{\delta}_{D} = \overline{\Delta}_{DP} ,
\eeq
the resonance is on-shell and the correlator contains a linear term in $t_{D}$,
\beq
 C_{{\mathrm{res}}}|_{\omega_{P} + \overline{\delta}_{D} = \overline{\Delta}_{DP}}
  = \frac{g_{BD^{\ast}} g_{\pi}}{f} \left ( \frac{1}{2} \right ) 
 \left ( \frac{\exponential^{-i \overline{\delta}_{D} t_{D} }}{2 v^{\prime}_{0}}  \right )  
 \left ( 
   \frac{\exponential^{-i \omega_{P} t_{P}}}{2 \omega_{P}} \right )
 \left ( \frac{i t_{D} }{ 2 v^{\prime}_{0}} \right ) ,
\eeq
which is an energy shift of the final state.  When
one takes the ratio of the $B\to D P$ correlator and the square root of the $D P \to D P$ correlator,
the $t_{D}$ dependence arising from the square brackets in Eq.~(\ref{eq:resonance_correlator}) (hence this linear
term in $t_{D}$),
is exactly cancelled by the contribution from the diagram 
in Fig.~\ref{fig:DPtoDP}.\footnote{This cancellation may not occur in partially-quenched QCD due to the
loss of unitarity~\cite{Bernard:1995ez,Lin:2002aj,Lin:2003tn}.}  The coupling $g_{\pi}$ is defined 
in Eq.~(\ref{eq:L1}), and $g_{BD^{\ast}}$ is a linear combination of the LEC's $\beta_{i}$ in
Eq.~(\ref{eq:chiral_O_expanded}).  Notice that $g_{BD^{\ast}}$ is different from $g_{BDP}$ and thus the
resonance contribution results in general in an additional unknown parameter for $B\to D P$ amplitude 
at the tree level.

\section{Beyond Tree Level}\label{sec:hardPi}

The use of tree-level \chpt\ is certainly limiting. While many systematic errors should cancel when looking at the ratio of $B\to DP$ to $B\to \overline{D}P$, going to higher order both in the chiral and heavy quark expansions is essential. Using the symmetry relations resulting from Eq.~(\ref{eq:chiral_O_expanded}), we could attempt an NLO calculation to make similar relationships at higher order, which is possible in the case of $K\to 2\pi$, as in, for example, Refs.~\cite{Laiho:2002jq,Laiho:2003uy,Aubin:2008vh}.

In order to treat these processes in the physical regime, we use the methods of Refs.~\cite{Flynn:2008tg,Bijnens:2009yr,Bijnens:2010ws}: Hard-pion \chpt\ (HP\chpt). As discussed earlier, this formalism uses the fact that one or more of the momenta in the final state very well may be hard, and at the physical point for $B\to DP$, this is true. For this section we will focus on $P=\pi$.

In order to apply HP\chpt\ to both $B\to D\pi$ and $B\to\Dbar\pi$, there are quite a few one-loop diagrams that we must evaluate. The result of a complete calculation (\emph{i.e.}, the sum of all one-loop diagrams) is expected to take the following generic form [working with the SU(2) chiral theory for now]
\be\label{eq:Moneloop}
	\cM = \cM^{\rm tree}\left[1 
	+ a\frac{m_\pi^2}{16\pi^2 f^2}
	\ln \left(\frac{m_\pi^2}{\Lambda^2}\right)
	+ L m_\pi^2
	\right] \ ,
\ee
where $\cM$ is one of the the particular amplitudes from Sec.~\ref{sec:ChPT_BtoDK}, and $\cM^{\rm tree}$ its tree-level value. $a$ is a coefficient that depends on the particular kinematics chosen for the diagram, and $L$ is a linear combination of low-energy constants as well as terms arising from higher-order chiral-level weak operators.\footnote{Repeating the spurion analysis of Sec.~\ref{sec:hl-chpt} would show in principle roughly 3-4 times as many LEC's arising at NLO relative to LO, but only certain combinations arise in calculations, \emph{eg}., Eq.~\ref{eq:Moneloop}, and as such there will effectively only be a small number of LEC's.} These would be determined from evaluating the full one-loop corrections to these amplitudes. We stress that $a$ and $L$ above will depend on all the hard
quantities, specifically the mass of the external $D$ meson and the
momenta of both the external $D$ meson and pion.  This dependence is
not known analytically, and it makes the LEC's non-universal when
varying the hard momenta.  However, at any fixed kinematics, the
values of the LEC's are still fixed\footnote{In practical lattice
calculations, one would have to vary the pion mass, and extrapolate to
the physical point.  In this procedure, it is inevitable to change the
momenta, and therefore the values of the LEC's.  Fortunately, since
the hard momenta are all much larger than the typical pion masses in
present and future lattice simulations, changes in the latter will
result in very small variations of the former.}.  Additionally, we
note that since all the hard scales are absorbed into the LEC's, we expect similar convergence as that of ordinary \chpt. Corrections to the heavy quark expansion will be more significant coming from the $D$-meson, and thus both $a$ and $L$ will have $\cO(1/M_D)$ corrections.

\begin{figure}[t]
\begin{center}
\includegraphics[width=3in]{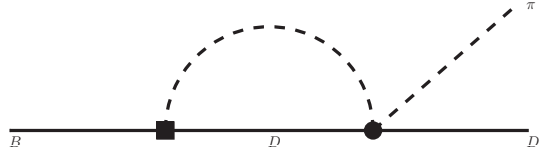}
\caption{One of the many one-loop diagrams that contribute to $B\to D\pi$, specifically one which shows the essential features that arise in HP\chpt.}
\label{fig:Sunset1_BtoDpi}
\end{center}
\end{figure}

In order to understand the specific details, we work through an example diagram, shown in Fig.~\ref{fig:Sunset1_BtoDpi}. To evaluate this diagram, we envision a lattice simulation where momentum will be conserved at the strong vertex, but need not be at the weak vertex. Thus, we define the momentum entering the weak vertex as $p_{\rm wk}$, the momentum flowing through the pion line is $\ell$ (the integration variable), and the external $D$ meson has velocity $v'$ and residual momentum $k$, so that this diagram has the form,
\bea
	\la D^{0} \pi^{-} | \op_{\chi,d} | B^{-}
	\ra_{\rm Fig.~\ref{fig:Sunset1_BtoDpi}}
	& = & 
	\frac{\la D^{0} \pi^{-} | \op_{\chi,d} |
	B^{-}\ra^{\rm tree}}{8f^2}
	\int
	\frac{d^d\ell}{(2\pi)^d}
	\frac{i}{\ell^2 - m_\pi^2 + i\epsilon}
	\frac{iv'\cdot(\ell-p_\pi)}{v'\cdot(\ell - k - 
	p_\pi)-\Delta+i\epsilon}\ ,
	\nonumber\\
	& \equiv & 
	\frac{\la D^{0} \pi^{-} | \op_{\chi,d} | B^{-}\ra^{\rm tree}}{8}
	I\ ,
	\label{eq:BtoDpi1loop}
\eea
where the coefficient arises from the weak vertex, and the momentum
injected into the weak vertex, $p_{\rm wk}$, is related to those
carried by the external $B$, $D$ and pion,
\be
	p_\pi + p_{D} = p_{B} + p_{\rm wk}\ .
\ee
$\Delta = M_D - M_B$ is the $D$-$B$ meson mass splitting (which is of order $1/m_c - 1/m_b$) and $k_B$ is the residual momentum of the $B$-meson.

This integral can be evaluated simply to obtain
\be\label{eq:HPint}
	I =
	\frac{1}{16\pi^2f^2}\left[
	\frac{v'\negcdot k+\Delta}{v'\negcdot (k+p_\pi)+\Delta+i\epsilon}I_2(m_\pi, v'\negcdot (k+p_\pi)+\Delta+i\epsilon)
	-m_\pi^2 \ln\left(\frac{m_\pi^2}{\Lambda^2}\right)\right]
	\ ,
\ee
with
\bea
	I_2(m,\delta)
	& = & -2\delta^2 \ln\left(\frac{m^2}{\Lambda^2}\right) - 4\delta^2 F(m/\delta) + 2\delta^2\ ,\\
	F(x) & = & \begin{cases}
	\sqrt{1-x^2}\tanh^{-1}\sqrt{1-x^2}\ , & 0\le x \le 1\\
	-\sqrt{x^2-1}\tan^{-1}\sqrt{x^2-1}\ , &  x \ge 1
	\end{cases}\ .
\eea

We can examine this case in the limit where $v'\negcdot k\gg m_\pi$, which is the hard-pion limit. In this limit we find
\be
	I_2(m_\pi, v'\negcdot (k+p_\pi)+\Delta) \approx
	-m_\pi^2\ln\left(\frac{m_\pi^2}{\Lambda^2}\right)\ ,
\ee
so that the full integral contributing to this diagram becomes either
\be
	I(p_\pi\approx 0) \to
	-2
	\frac{m_\pi^2}{16\pi^2f^2}
	\ln\left(\frac{m_\pi^2}{\Lambda^2}\right)
	\ ,
\ee
if we insert momentum into the weak vertex such that $p_\pi\approx 0$, or 
\be
	I(p_\pi\approx k)\to-
	\frac32\frac{m_\pi^2}{16\pi^2f^2}
	\ln\left(\frac{m_\pi^2}{\Lambda^2}\right)
	\ ,
\ee
if we choose $p_{\rm wk}$ such that $p_\pi\approx k$. These would give rise to different values of the coefficient $a$ in Eq.~(\ref{eq:Moneloop}).  This can thus be extended to all of the diagrams that would contribute to one-loop order, and for each chosen set of kinematics, we would be able to find different expressions for $a$ in Eq.~(\ref{eq:Moneloop}), and in general, the LEC $L$ in that equation would have an unknown dependence on the kinematics. However, the pion mass dependence is well determined.

We close this section by noting that HP\chpt\ is not applicable for extracting the strong phases of $B$ decays via the computation of the one-loop diagram in Fig.~\ref{fig:Sunset1_BtoDpi}.   The imaginary part in this diagram is proportional to $\sqrt{[v^{\prime}\cdot (p_{\pi}+k)]^{2} - m^{2}_{\pi}}$, therefore grows with the increasing momenta carried by the final-state mesons, leading to the failure of the chiral expansion when $p_{D}$ and $p_{\pi}$ are large.  This can be understood by noting that the imaginary part arises from the contribution in which both mesons in the loop are on-shell, and therefore cannot be soft.  

\section{Summary and Outlook}\label{sec:disc}
In this paper, we proposed a strategy for calculating $B\to D P$ and $B\to \Dbar P$ ($P$ is a Goldstone
boson) decay amplitudes via lattice calculations. Indeed the real part is accessible directly via four-point function calculations on the lattice as it does not suffer from the Maiani-Testa No-Go Theorem, though it is computationally demanding. As an approximation, one can invoke the chiral expansion, specifically taking into account the large momenta of the final state mesons. We argue that this hard-pion chiral expansion is valid for these decays, for similar reasons to those in
semileptonic $B$ decays and in $K\to \pi\pi$ amplitudes. In general, this hard-pion chiral expansion results in
momentum dependence of low-energy constants and the coefficients of the chiral logarithms. From our investigation
of the structure of a typical one-loop diagram (Fig.~\ref{fig:Sunset1_BtoDpi}), it is shown explicitly how this
occurs for the $B\to D P$ amplitudes.
 
We constructed the leading-order operators, relevant to these decays, in the chiral effective theory. We studied the
tree-level resonance contributions in the framework of HM\chpt, and showed that these contributions are accompanied
by combinations of the LEC's which are different from that for the corresponding leading-order $B\to D P$ and $B\to \Dbar P$
amplitudes. As such, incorporating resonances in the study of the lattice correlators allows us to extract some of the LEC's that
are not accessible by applying \chpt\ naively.
 
To complete this inital approach, the complete one-loop contributions must be calculated~\cite{Aubin_Lin_Soni:in_progress}. One can
combine these \chpt\ results with lattice simulations to compute the real parts of $B\to D P$ and $B\to \Dbar P$
decay amplitudes. Although the lattice calculation for the imaginary parts of these matrix elements is
challenging, their real parts can already provide important information for an accurate determination of
the angle $\gamma$ in the $b{-}d$ unitarity triangle in the CKM matrix. The real part gives a lower bound to the absolute value of the amplitude, which would be very useful in the phenomenology of $\gamma$-extraction, and by combining this with the information on strong phases from $B$ and $D$ decays, the absolute magnitude of the amplitude can also be deduced.
 
Finally, let us note that, in the long run, as the lattice program succeeds in evaluating $r^{\rm red}_{BP}$ and with experimental studies using larger data samples, experiment will be able to pin down $r_{BP}$ with increasing precision.  We envision that a combination of these efforts could lead to an improvement in determinations of $\gamma$ to about 10\% in 3-5 years. In the longer term, with the use of even more powerful computers and with data from Super-LHCb and Super-$B$ factories, the error could be reduced to a few percent. These improved determinations should allow a useful constraint on $V^{\rm combo}_{CKM}$ [Eq.~(\ref{eq:reduced_r})] and consequently on $V_{ub}$ since all the other factors therein are already known quite well. Given the serious difficulties \cite{Lunghi:2010gv} in a precise determination of $V_{ub}$ through the conventional semileptonic methods, having an independent constraint via purely hadronic decays: $B\to D(\Dbar)P$ may also prove useful.

\acknowledgments
This work has been supported by the US DOE under grant number DE-AC02-98CH10886 and Taiwanese NSC grant 99-2112-M-009-004-MY3. C.-J.D.L. thanks the hospitality of Fordham University and Brookhaven National Lab during the progress of this work.

\bibliography{refs}

\begin{thebibliography}{56}
\expandafter\ifx\csname natexlab\endcsname\relax\def\natexlab#1{#1}\fi
\expandafter\ifx\csname bibnamefont\endcsname\relax
  \def\bibnamefont#1{#1}\fi
\expandafter\ifx\csname bibfnamefont\endcsname\relax
  \def\bibfnamefont#1{#1}\fi
\expandafter\ifx\csname citenamefont\endcsname\relax
  \def\citenamefont#1{#1}\fi
\expandafter\ifx\csname url\endcsname\relax
  \def\url#1{\texttt{#1}}\fi
\expandafter\ifx\csname urlprefix\endcsname\relax\def\urlprefix{URL }\fi
\providecommand{\bibinfo}[2]{#2}
\providecommand{\eprint}[2][]{\url{#2}}

\bibitem[{\citenamefont{Asner et~al.}(2010)}]{Asner_rev10}
\bibinfo{author}{\bibfnamefont{D.}~\bibnamefont{Asner}} \bibnamefont{et~al.}
  (\bibinfo{collaboration}{Heavy Flavor Averaging Group})
  (\bibinfo{year}{2010}), \eprint{arXiv:1010.1589}.

\bibitem[{\citenamefont{Browder et~al.}(2009)\citenamefont{Browder, Gershon,
  Pirjol, Soni, and Zupan}}]{RMP09}
\bibinfo{author}{\bibfnamefont{T.~E.} \bibnamefont{Browder}},
  \bibinfo{author}{\bibfnamefont{T.}~\bibnamefont{Gershon}},
  \bibinfo{author}{\bibfnamefont{D.}~\bibnamefont{Pirjol}},
  \bibinfo{author}{\bibfnamefont{A.}~\bibnamefont{Soni}}, \bibnamefont{and}
  \bibinfo{author}{\bibfnamefont{J.}~\bibnamefont{Zupan}},
  \bibinfo{journal}{Rev.Mod.Phys.} \textbf{\bibinfo{volume}{81}},
  \bibinfo{pages}{1887} (\bibinfo{year}{2009}), \eprint{arXiv:0802.3201}.

\bibitem[{\citenamefont{Gronau and London.}(1991)}]{Gronau:1990ra}
\bibinfo{author}{\bibfnamefont{M.}~\bibnamefont{Gronau}} \bibnamefont{and}
  \bibinfo{author}{\bibfnamefont{D.}~\bibnamefont{London.}},
  \bibinfo{journal}{Phys. Lett.} \textbf{\bibinfo{volume}{B253}},
  \bibinfo{pages}{483} (\bibinfo{year}{1991}).

\bibitem[{\citenamefont{Gronau and Wyler}(1991)}]{Gronau:1991dp}
\bibinfo{author}{\bibfnamefont{M.}~\bibnamefont{Gronau}} \bibnamefont{and}
  \bibinfo{author}{\bibfnamefont{D.}~\bibnamefont{Wyler}},
  \bibinfo{journal}{Phys. Lett.} \textbf{\bibinfo{volume}{B265}},
  \bibinfo{pages}{172} (\bibinfo{year}{1991}).

\bibitem[{\citenamefont{Atwood et~al.}(1997)\citenamefont{Atwood, Dunietz, and
  Soni}}]{ADS96}
\bibinfo{author}{\bibfnamefont{D.}~\bibnamefont{Atwood}},
  \bibinfo{author}{\bibfnamefont{I.}~\bibnamefont{Dunietz}}, \bibnamefont{and}
  \bibinfo{author}{\bibfnamefont{A.}~\bibnamefont{Soni}},
  \bibinfo{journal}{Phys.Rev.Lett.} \textbf{\bibinfo{volume}{78}},
  \bibinfo{pages}{3257} (\bibinfo{year}{1997}), \eprint{hep-ph/9612433}.

\bibitem[{\citenamefont{Atwood et~al.}(2001)\citenamefont{Atwood, Dunietz, and
  Soni}}]{ADS00}
\bibinfo{author}{\bibfnamefont{D.}~\bibnamefont{Atwood}},
  \bibinfo{author}{\bibfnamefont{I.}~\bibnamefont{Dunietz}}, \bibnamefont{and}
  \bibinfo{author}{\bibfnamefont{A.}~\bibnamefont{Soni}},
  \bibinfo{journal}{Phys.Rev.} \textbf{\bibinfo{volume}{D63}},
  \bibinfo{pages}{036005} (\bibinfo{year}{2001}), \eprint{hep-ph/0008090}.

\bibitem[{\citenamefont{Giri et~al.}(2003)\citenamefont{Giri, Grossman, Soffer,
  and Zupan}}]{GGSZ}
\bibinfo{author}{\bibfnamefont{A.}~\bibnamefont{Giri}},
  \bibinfo{author}{\bibfnamefont{Y.}~\bibnamefont{Grossman}},
  \bibinfo{author}{\bibfnamefont{A.}~\bibnamefont{Soffer}}, \bibnamefont{and}
  \bibinfo{author}{\bibfnamefont{J.}~\bibnamefont{Zupan}},
  \bibinfo{journal}{Phys.Rev.} \textbf{\bibinfo{volume}{D68}},
  \bibinfo{pages}{054018} (\bibinfo{year}{2003}), \eprint{hep-ph/0303187}.

\bibitem[{\citenamefont{Soffer}(1998)}]{Abi98}
\bibinfo{author}{\bibfnamefont{A.}~\bibnamefont{Soffer}}
  (\bibinfo{year}{1998}), \eprint{hep-ex/9801018}.

\bibitem[{\citenamefont{Silva and Soffer}(2000)}]{SS99}
\bibinfo{author}{\bibfnamefont{J.~P.} \bibnamefont{Silva}} \bibnamefont{and}
  \bibinfo{author}{\bibfnamefont{A.}~\bibnamefont{Soffer}},
  \bibinfo{journal}{Phys.Rev.} \textbf{\bibinfo{volume}{D61}},
  \bibinfo{pages}{112001} (\bibinfo{year}{2000}), \eprint{hep-ph/9912242}.

\bibitem[{\citenamefont{Atwood and Soni}(2003)}]{Atwood:2003mj}
\bibinfo{author}{\bibfnamefont{D.}~\bibnamefont{Atwood}} \bibnamefont{and}
  \bibinfo{author}{\bibfnamefont{A.}~\bibnamefont{Soni}},
  \bibinfo{journal}{Phys.Rev.} \textbf{\bibinfo{volume}{D68}},
  \bibinfo{pages}{033003} (\bibinfo{year}{2003}), \eprint{hep-ph/0304085}.

\bibitem[{\citenamefont{Rosner}(2011)}]{Rosner:2011sj}
\bibinfo{author}{\bibfnamefont{J.~L.} \bibnamefont{Rosner}}
  (\bibinfo{year}{2011}), \eprint{arXiv:1107.2023}.

\bibitem[{\citenamefont{Thomas}(2010)}]{Thomas:2010zzj}
\bibinfo{author}{\bibfnamefont{C.}~\bibnamefont{Thomas}}
  (\bibinfo{collaboration}{CLEO-c Collaboration}), \bibinfo{journal}{PoS}
  \textbf{\bibinfo{volume}{FPCP2010}}, \bibinfo{pages}{008}
  (\bibinfo{year}{2010}).

\bibitem[{\citenamefont{Malde and Wilkinson}(2011)}]{Malde:2011mk}
\bibinfo{author}{\bibfnamefont{S.}~\bibnamefont{Malde}} \bibnamefont{and}
  \bibinfo{author}{\bibfnamefont{G.}~\bibnamefont{Wilkinson}},
  \bibinfo{journal}{Phys.Lett.} \textbf{\bibinfo{volume}{B701}},
  \bibinfo{pages}{353} (\bibinfo{year}{2011}), \eprint{arXiv:1104.2731}.

\bibitem[{\citenamefont{Ricciardi}(2011)}]{Ricciardi:2011vk}
\bibinfo{author}{\bibfnamefont{S.}~\bibnamefont{Ricciardi}}
  (\bibinfo{year}{2011}), \eprint{arXiv:1101.4855}.

\bibitem[{\citenamefont{Poluektov}(2007)}]{AP07}
\bibinfo{author}{\bibfnamefont{A.}~\bibnamefont{Poluektov}}
  (\bibinfo{year}{2007}), \bibinfo{note}{ph.D. Thesis}.

\bibitem[{\citenamefont{Bernard et~al.}(1990)\citenamefont{Bernard, Simone, and
  Soni}}]{Simone1}
\bibinfo{author}{\bibfnamefont{C.~W.} \bibnamefont{Bernard}},
  \bibinfo{author}{\bibfnamefont{J.}~\bibnamefont{Simone}}, \bibnamefont{and}
  \bibinfo{author}{\bibfnamefont{A.}~\bibnamefont{Soni}},
  \bibinfo{journal}{Nucl.Phys.Proc.Suppl.} \textbf{\bibinfo{volume}{17}},
  \bibinfo{pages}{504} (\bibinfo{year}{1990}).

\bibitem[{\citenamefont{Bernard et~al.}(1991)\citenamefont{Bernard, Simone, and
  Soni}}]{Simone2}
\bibinfo{author}{\bibfnamefont{C.~W.} \bibnamefont{Bernard}},
  \bibinfo{author}{\bibfnamefont{J.~N.} \bibnamefont{Simone}},
  \bibnamefont{and} \bibinfo{author}{\bibfnamefont{A.}~\bibnamefont{Soni}},
  \bibinfo{journal}{Nucl.Phys.Proc.Suppl.} \textbf{\bibinfo{volume}{20}},
  \bibinfo{pages}{434} (\bibinfo{year}{1991}).

\bibitem[{\citenamefont{Abada et~al.}(1990)}]{MartiDKP}
\bibinfo{author}{\bibfnamefont{A.}~\bibnamefont{Abada}} \bibnamefont{et~al.}
  (\bibinfo{collaboration}{European Lattice Collaboration}),
  \bibinfo{journal}{Nucl.Phys.Proc.Suppl.} \textbf{\bibinfo{volume}{17}},
  \bibinfo{pages}{518} (\bibinfo{year}{1990}).

\bibitem[{\citenamefont{Maiani and Testa}(1990)}]{Maiani:1990ca}
\bibinfo{author}{\bibfnamefont{L.}~\bibnamefont{Maiani}} \bibnamefont{and}
  \bibinfo{author}{\bibfnamefont{M.}~\bibnamefont{Testa}},
  \bibinfo{journal}{Phys. Lett.} \textbf{\bibinfo{volume}{B245}},
  \bibinfo{pages}{585} (\bibinfo{year}{1990}).

\bibitem[{\citenamefont{Lin et~al.}(2001)\citenamefont{Lin, Martinelli,
  Sachrajda, and Testa}}]{Lin:2001ek}
\bibinfo{author}{\bibfnamefont{C.}~\bibnamefont{Lin}},
  \bibinfo{author}{\bibfnamefont{G.}~\bibnamefont{Martinelli}},
  \bibinfo{author}{\bibfnamefont{C.~T.} \bibnamefont{Sachrajda}},
  \bibnamefont{and} \bibinfo{author}{\bibfnamefont{M.}~\bibnamefont{Testa}},
  \bibinfo{journal}{Nucl.Phys.} \textbf{\bibinfo{volume}{B619}},
  \bibinfo{pages}{467} (\bibinfo{year}{2001}), \eprint{hep-lat/0104006}.

\bibitem[{\citenamefont{Lellouch and Luscher}(2001)}]{Lellouch:2000pv}
\bibinfo{author}{\bibfnamefont{L.}~\bibnamefont{Lellouch}} \bibnamefont{and}
  \bibinfo{author}{\bibfnamefont{M.}~\bibnamefont{Luscher}},
  \bibinfo{journal}{Commun. Math. Phys.} \textbf{\bibinfo{volume}{219}},
  \bibinfo{pages}{31} (\bibinfo{year}{2001}), \eprint{hep-lat/0003023}.

\bibitem[{\citenamefont{Blum et~al.}(2011{\natexlab{a}})\citenamefont{Blum,
  Boyle, Christ, Garron, Goode et~al.}}]{Qi_ReA0_11}
\bibinfo{author}{\bibfnamefont{T.}~\bibnamefont{Blum}},
  \bibinfo{author}{\bibfnamefont{P.}~\bibnamefont{Boyle}},
  \bibinfo{author}{\bibfnamefont{N.}~\bibnamefont{Christ}},
  \bibinfo{author}{\bibfnamefont{N.}~\bibnamefont{Garron}},
  \bibinfo{author}{\bibfnamefont{E.}~\bibnamefont{Goode}}, \bibnamefont{et~al.}
  (\bibinfo{year}{2011}{\natexlab{a}}), \eprint{arXiv:1106.2714}.

\bibitem[{\citenamefont{Blum et~al.}(2011{\natexlab{b}})\citenamefont{Blum,
  Boyle, Christ, Garron, Goode et~al.}}]{Blum:2011ng}
\bibinfo{author}{\bibfnamefont{T.}~\bibnamefont{Blum}},
  \bibinfo{author}{\bibfnamefont{P.}~\bibnamefont{Boyle}},
  \bibinfo{author}{\bibfnamefont{N.}~\bibnamefont{Christ}},
  \bibinfo{author}{\bibfnamefont{N.}~\bibnamefont{Garron}},
  \bibinfo{author}{\bibfnamefont{E.}~\bibnamefont{Goode}}, \bibnamefont{et~al.}
  (\bibinfo{year}{2011}{\natexlab{b}}), \eprint{arXiv:1111.1699}.

\bibitem[{\citenamefont{Wise}(1992)}]{Wise:1992hn}
\bibinfo{author}{\bibfnamefont{M.~B.} \bibnamefont{Wise}},
  \bibinfo{journal}{Phys. Rev.} \textbf{\bibinfo{volume}{D45}},
  \bibinfo{pages}{2188} (\bibinfo{year}{1992}).

\bibitem[{\citenamefont{Burdman and Donoghue}(1992)}]{Burdman:1992gh}
\bibinfo{author}{\bibfnamefont{G.}~\bibnamefont{Burdman}} \bibnamefont{and}
  \bibinfo{author}{\bibfnamefont{J.~F.} \bibnamefont{Donoghue}},
  \bibinfo{journal}{Phys. Lett.} \textbf{\bibinfo{volume}{B280}},
  \bibinfo{pages}{287} (\bibinfo{year}{1992}).

\bibitem[{\citenamefont{Yan et~al.}(1992)}]{Yan:1992gz}
\bibinfo{author}{\bibfnamefont{T.-M.} \bibnamefont{Yan}} \bibnamefont{et~al.},
  \bibinfo{journal}{Phys. Rev.} \textbf{\bibinfo{volume}{D46}},
  \bibinfo{pages}{1148} (\bibinfo{year}{1992}), \bibinfo{note}{[Erratum-ibid
  {\bf D55}, 5851 (1997)].}

\bibitem[{\citenamefont{Cho}(1992)}]{Cho:1992gg}
\bibinfo{author}{\bibfnamefont{P.~L.} \bibnamefont{Cho}},
  \bibinfo{journal}{Phys. Lett.} \textbf{\bibinfo{volume}{B285}},
  \bibinfo{pages}{145} (\bibinfo{year}{1992}), \eprint{hep-ph/9203225}.

\bibitem[{\citenamefont{Cho}(1993)}]{Cho:1992cf}
\bibinfo{author}{\bibfnamefont{P.~L.} \bibnamefont{Cho}},
  \bibinfo{journal}{Nucl. Phys.} \textbf{\bibinfo{volume}{B396}},
  \bibinfo{pages}{183} (\bibinfo{year}{1993}), \bibinfo{note}{[Erratum-ibid.
  {\bf B421}, 683 (1994)]}, \eprint{hep-ph/9208244}.

\bibitem[{\citenamefont{Manohar and Wise}(2000)}]{Manohar:2000dt}
\bibinfo{author}{\bibfnamefont{A.~V.} \bibnamefont{Manohar}} \bibnamefont{and}
  \bibinfo{author}{\bibfnamefont{M.~B.} \bibnamefont{Wise}},
  \bibinfo{journal}{Camb. Monogr. Part. Phys. Nucl. Phys. Cosmol.}
  \textbf{\bibinfo{volume}{10}}, \bibinfo{pages}{1} (\bibinfo{year}{2000}).

\bibitem[{\citenamefont{Antonio et~al.}(2008)}]{Antonio07}
\bibinfo{author}{\bibfnamefont{D.}~\bibnamefont{Antonio}} \bibnamefont{et~al.}
  (\bibinfo{collaboration}{RBC Collaboration, UKQCD Collaboration}),
  \bibinfo{journal}{Phys.Rev.Lett.} \textbf{\bibinfo{volume}{100}},
  \bibinfo{pages}{032001} (\bibinfo{year}{2008}), \eprint{hep-ph/0702042}.

\bibitem[{\citenamefont{Allton et~al.}(2008)}]{Allton08}
\bibinfo{author}{\bibfnamefont{C.}~\bibnamefont{Allton}} \bibnamefont{et~al.}
  (\bibinfo{collaboration}{RBC-UKQCD Collaboration}),
  \bibinfo{journal}{Phys.Rev.} \textbf{\bibinfo{volume}{D78}},
  \bibinfo{pages}{114509} (\bibinfo{year}{2008}), \eprint{arXiv:0804.0473}.

\bibitem[{\citenamefont{Flynn and Sachrajda}(2009)}]{Flynn:2008tg}
\bibinfo{author}{\bibfnamefont{J.}~\bibnamefont{Flynn}} \bibnamefont{and}
  \bibinfo{author}{\bibfnamefont{C.}~\bibnamefont{Sachrajda}}
  (\bibinfo{collaboration}{RBC Collaboration, UKQCD Collaboration}),
  \bibinfo{journal}{Nucl.Phys.} \textbf{\bibinfo{volume}{B812}},
  \bibinfo{pages}{64} (\bibinfo{year}{2009}), \eprint{arXiv:0809.1229}.

\bibitem[{\citenamefont{Bijnens and Celis}(2009)}]{Bijnens:2009yr}
\bibinfo{author}{\bibfnamefont{J.}~\bibnamefont{Bijnens}} \bibnamefont{and}
  \bibinfo{author}{\bibfnamefont{A.}~\bibnamefont{Celis}},
  \bibinfo{journal}{Phys.Lett.} \textbf{\bibinfo{volume}{B680}},
  \bibinfo{pages}{466} (\bibinfo{year}{2009}), \eprint{arXiv:0906.0302}.

\bibitem[{\citenamefont{Bijnens and Jemos}(2010)}]{Bijnens:2010ws}
\bibinfo{author}{\bibfnamefont{J.}~\bibnamefont{Bijnens}} \bibnamefont{and}
  \bibinfo{author}{\bibfnamefont{I.}~\bibnamefont{Jemos}},
  \bibinfo{journal}{Nucl.Phys.} \textbf{\bibinfo{volume}{B840}},
  \bibinfo{pages}{54} (\bibinfo{year}{2010}), \eprint{arXiv:1006.1197}.

\bibitem[{\citenamefont{Colangelo}(2011)}]{ColangeloLAT11}
\bibinfo{author}{\bibfnamefont{G.}~\bibnamefont{Colangelo}}
  (\bibinfo{year}{2011}), \bibinfo{note}{talk given at Lattice 2011}.

\bibitem[{\citenamefont{Gasser and Leutwyler}(1984)}]{Gasser:1983yg}
\bibinfo{author}{\bibfnamefont{J.}~\bibnamefont{Gasser}} \bibnamefont{and}
  \bibinfo{author}{\bibfnamefont{H.}~\bibnamefont{Leutwyler}},
  \bibinfo{journal}{Ann. Phys.} \textbf{\bibinfo{volume}{158}},
  \bibinfo{pages}{142} (\bibinfo{year}{1984}).

\bibitem[{\citenamefont{Gasser and Leutwyler}(1985)}]{Gasser:1984gg}
\bibinfo{author}{\bibfnamefont{J.}~\bibnamefont{Gasser}} \bibnamefont{and}
  \bibinfo{author}{\bibfnamefont{H.}~\bibnamefont{Leutwyler}},
  \bibinfo{journal}{Nucl. Phys.} \textbf{\bibinfo{volume}{B250}},
  \bibinfo{pages}{465} (\bibinfo{year}{1985}).

\bibitem[{\citenamefont{Aubin and Bernard}(2006)}]{Aubin:2005aq}
\bibinfo{author}{\bibfnamefont{C.}~\bibnamefont{Aubin}} \bibnamefont{and}
  \bibinfo{author}{\bibfnamefont{C.}~\bibnamefont{Bernard}},
  \bibinfo{journal}{Phys. Rev.} \textbf{\bibinfo{volume}{D73}},
  \bibinfo{pages}{014515} (\bibinfo{year}{2006}), \eprint{hep-lat/0510088}.

\bibitem[{\citenamefont{Grinstein et~al.}(1992)\citenamefont{Grinstein,
  Jenkins, Manohar, Savage, and Wise}}]{Grinstein:1992qt}
\bibinfo{author}{\bibfnamefont{B.}~\bibnamefont{Grinstein}},
  \bibinfo{author}{\bibfnamefont{E.~E.} \bibnamefont{Jenkins}},
  \bibinfo{author}{\bibfnamefont{A.~V.} \bibnamefont{Manohar}},
  \bibinfo{author}{\bibfnamefont{M.~J.} \bibnamefont{Savage}},
  \bibnamefont{and} \bibinfo{author}{\bibfnamefont{M.~B.} \bibnamefont{Wise}},
  \bibinfo{journal}{Nucl. Phys.} \textbf{\bibinfo{volume}{B380}},
  \bibinfo{pages}{369} (\bibinfo{year}{1992}), \eprint{hep-ph/9204207}.

\bibitem[{\citenamefont{Ohki et~al.}(2008)\citenamefont{Ohki, Matsufuru, and
  Onogi}}]{Ohki:2008py}
\bibinfo{author}{\bibfnamefont{H.}~\bibnamefont{Ohki}},
  \bibinfo{author}{\bibfnamefont{H.}~\bibnamefont{Matsufuru}},
  \bibnamefont{and} \bibinfo{author}{\bibfnamefont{T.}~\bibnamefont{Onogi}},
  \bibinfo{journal}{Phys. Rev.} \textbf{\bibinfo{volume}{D77}},
  \bibinfo{pages}{094509} (\bibinfo{year}{2008}), \eprint{arXiv:0802.1563}.

\bibitem[{\citenamefont{Becirevic et~al.}(2009)\citenamefont{Becirevic,
  Blossier, Chang, and Haas}}]{Becirevic:2009yb}
\bibinfo{author}{\bibfnamefont{D.}~\bibnamefont{Becirevic}},
  \bibinfo{author}{\bibfnamefont{B.}~\bibnamefont{Blossier}},
  \bibinfo{author}{\bibfnamefont{E.}~\bibnamefont{Chang}}, \bibnamefont{and}
  \bibinfo{author}{\bibfnamefont{B.}~\bibnamefont{Haas}},
  \bibinfo{journal}{Phys. Lett.} \textbf{\bibinfo{volume}{B679}},
  \bibinfo{pages}{231} (\bibinfo{year}{2009}), \eprint{arXiv:0905.3355}.

\bibitem[{\citenamefont{{W.~Detmold, C.-J.~D.~Lin, and
  S.~Meinel}}()}]{Detmold:2011bp}
\bibinfo{author}{\bibnamefont{{W.~Detmold, C.-J.~D.~Lin, and S.~Meinel}}},
  \bibinfo{note}{{arXiv:1109.2480}}.

\bibitem[{\citenamefont{Boyd and Grinstein}(1995)}]{Boyd:1995pa}
\bibinfo{author}{\bibfnamefont{C.~G.} \bibnamefont{Boyd}} \bibnamefont{and}
  \bibinfo{author}{\bibfnamefont{B.}~\bibnamefont{Grinstein}},
  \bibinfo{journal}{Nucl. Phys.} \textbf{\bibinfo{volume}{B442}},
  \bibinfo{pages}{205} (\bibinfo{year}{1995}), \eprint{hep-ph/9402340}.

\bibitem[{\citenamefont{Stewart}(1998)}]{Stewart:1998ke}
\bibinfo{author}{\bibfnamefont{I.~W.} \bibnamefont{Stewart}},
  \bibinfo{journal}{Nucl. Phys.} \textbf{\bibinfo{volume}{B529}},
  \bibinfo{pages}{62} (\bibinfo{year}{1998}), \eprint{hep-ph/9803227}.

\bibitem[{\citenamefont{Bijnens and Jemos}(2011)}]{Bijnens:2010jg}
\bibinfo{author}{\bibfnamefont{J.}~\bibnamefont{Bijnens}} \bibnamefont{and}
  \bibinfo{author}{\bibfnamefont{I.}~\bibnamefont{Jemos}},
  \bibinfo{journal}{Nucl.Phys.} \textbf{\bibinfo{volume}{B846}},
  \bibinfo{pages}{145} (\bibinfo{year}{2011}), \eprint{arXiv:1011.6531}.

\bibitem[{\citenamefont{Detmold and Lin}(2007)}]{Detmold:2006gh}
\bibinfo{author}{\bibfnamefont{W.}~\bibnamefont{Detmold}} \bibnamefont{and}
  \bibinfo{author}{\bibfnamefont{C.-J.~D.} \bibnamefont{Lin}},
  \bibinfo{journal}{Phys. Rev.} \textbf{\bibinfo{volume}{D76}},
  \bibinfo{pages}{014501} (\bibinfo{year}{2007}), \eprint{hep-lat/0612028}.

\bibitem[{\citenamefont{Becirevic et~al.}(2007)\citenamefont{Becirevic, Fajfer,
  and Kamenik}}]{Becirevic:2007dg}
\bibinfo{author}{\bibfnamefont{D.}~\bibnamefont{Becirevic}},
  \bibinfo{author}{\bibfnamefont{S.}~\bibnamefont{Fajfer}}, \bibnamefont{and}
  \bibinfo{author}{\bibfnamefont{J.}~\bibnamefont{Kamenik}}
  (\bibinfo{year}{2007}), \eprint{arXiv:0710.3496 [hep-lat]}.

\bibitem[{\citenamefont{Aglietti et~al.}(1992)\citenamefont{Aglietti,
  Crisafulli, and Masetti}}]{Aglietti:1992in}
\bibinfo{author}{\bibfnamefont{U.}~\bibnamefont{Aglietti}},
  \bibinfo{author}{\bibfnamefont{M.}~\bibnamefont{Crisafulli}},
  \bibnamefont{and} \bibinfo{author}{\bibfnamefont{M.}~\bibnamefont{Masetti}},
  \bibinfo{journal}{Phys.Lett.} \textbf{\bibinfo{volume}{B294}},
  \bibinfo{pages}{281} (\bibinfo{year}{1992}).

\bibitem[{\citenamefont{Bernard and Golterman}(1996)}]{Bernard:1995ez}
\bibinfo{author}{\bibfnamefont{C.~W.} \bibnamefont{Bernard}} \bibnamefont{and}
  \bibinfo{author}{\bibfnamefont{M.~F.} \bibnamefont{Golterman}},
  \bibinfo{journal}{Phys.Rev.} \textbf{\bibinfo{volume}{D53}},
  \bibinfo{pages}{476} (\bibinfo{year}{1996}), \eprint{hep-lat/9507004}.

\bibitem[{\citenamefont{{C.-J.~D.~Lin, G.~Martinelli, E.~Pallante,
  C.~T.~Sachrajda, and G.~Villadoro}}(2003)}]{Lin:2002aj}
\bibinfo{author}{\bibnamefont{{C.-J.~D.~Lin, G.~Martinelli, E.~Pallante,
  C.~T.~Sachrajda, and G.~Villadoro}}}, \bibinfo{journal}{Phys.Lett.}
  \textbf{\bibinfo{volume}{B553}}, \bibinfo{pages}{229} (\bibinfo{year}{2003}),
  \eprint{hep-lat/0211043}.

\bibitem[{\citenamefont{Lin et~al.}(2004)\citenamefont{Lin, Martinelli,
  Pallante, Sachrajda, and Villadoro}}]{Lin:2003tn}
\bibinfo{author}{\bibfnamefont{C.-J.~D.} \bibnamefont{Lin}},
  \bibinfo{author}{\bibfnamefont{G.}~\bibnamefont{Martinelli}},
  \bibinfo{author}{\bibfnamefont{E.}~\bibnamefont{Pallante}},
  \bibinfo{author}{\bibfnamefont{C.~T.} \bibnamefont{Sachrajda}},
  \bibnamefont{and}
  \bibinfo{author}{\bibfnamefont{G.}~\bibnamefont{Villadoro}},
  \bibinfo{journal}{Phys. Lett.} \textbf{\bibinfo{volume}{B581}},
  \bibinfo{pages}{207} (\bibinfo{year}{2004}), \eprint{hep-lat/0308014}.

\bibitem[{\citenamefont{Laiho and Soni}(2002)}]{Laiho:2002jq}
\bibinfo{author}{\bibfnamefont{J.}~\bibnamefont{Laiho}} \bibnamefont{and}
  \bibinfo{author}{\bibfnamefont{A.}~\bibnamefont{Soni}},
  \bibinfo{journal}{Phys. Rev.} \textbf{\bibinfo{volume}{D65}},
  \bibinfo{pages}{114020} (\bibinfo{year}{2002}), \eprint{hep-ph/0203106}.

\bibitem[{\citenamefont{Laiho and Soni}(2005)}]{Laiho:2003uy}
\bibinfo{author}{\bibfnamefont{J.}~\bibnamefont{Laiho}} \bibnamefont{and}
  \bibinfo{author}{\bibfnamefont{A.}~\bibnamefont{Soni}},
  \bibinfo{journal}{Phys. Rev.} \textbf{\bibinfo{volume}{D71}},
  \bibinfo{pages}{014021} (\bibinfo{year}{2005}), \eprint{hep-lat/0306035}.

\bibitem[{\citenamefont{Aubin et~al.}(2008)\citenamefont{Aubin, Laiho, Li, and
  Lin}}]{Aubin:2008vh}
\bibinfo{author}{\bibfnamefont{C.}~\bibnamefont{Aubin}},
  \bibinfo{author}{\bibfnamefont{J.}~\bibnamefont{Laiho}},
  \bibinfo{author}{\bibfnamefont{S.}~\bibnamefont{Li}}, \bibnamefont{and}
  \bibinfo{author}{\bibfnamefont{M.~F.} \bibnamefont{Lin}},
  \bibinfo{journal}{Phys. Rev.} \textbf{\bibinfo{volume}{D78}},
  \bibinfo{pages}{094505} (\bibinfo{year}{2008}), \eprint{arXiv:0808.3264}.

\bibitem[{\citenamefont{{C.~Aubin, C.-J.~D.~Lin,
  A.~Soni}}()}]{Aubin_Lin_Soni:in_progress}
\bibinfo{author}{\bibnamefont{{C.~Aubin, C.-J.~D.~Lin, A.~Soni}}},
  \bibinfo{note}{{work in progress}}.

\bibitem[{\citenamefont{Lunghi and Soni}(2011)}]{Lunghi:2010gv}
\bibinfo{author}{\bibfnamefont{E.}~\bibnamefont{Lunghi}} \bibnamefont{and}
  \bibinfo{author}{\bibfnamefont{A.}~\bibnamefont{Soni}},
  \bibinfo{journal}{Phys.Lett.} \textbf{\bibinfo{volume}{B697}},
  \bibinfo{pages}{323} (\bibinfo{year}{2011}), \eprint{arXiv:1010.6069}.

\end{thebibliography}

\end{document}